\documentclass[aps,prb,amsmath,superscriptaddress,twocolumn,bibnotes,longbibliography,floatfix]{revtex4-2}

\usepackage{amsfonts}
\usepackage{graphicx}
\usepackage{color}
\usepackage{dsfont}
\usepackage{verbatim}
\usepackage{mathtools}
\usepackage[normalem]{ulem}
\usepackage{comment}
\usepackage{stackrel}
\usepackage{bm}

\usepackage[colorlinks=true, 
linkcolor=blue, 
urlcolor=blue,
citecolor=blue]{hyperref}
\usepackage{orcidlink}
\usepackage[linesnumbered]{algorithm2e}

\RestyleAlgo{ruled}
\SetKwComment{Comment}{/* }{ */}

\DeclareMathOperator{\sgn}{sgn}

\definecolor{myblue}{rgb}{.93, .93, 1}

\newcommand{\bsub}{\begin{subequations}}
	\newcommand{\esub}{\end{subequations}}

\begin{document}
	
	\title{Monte Carlo solver and renormalization of Migdal-Eliashberg spin chain}
	
	\author{Yang-Zhi~Chou~\orcidlink{0000-0001-7955-0918}}\email{yzchou@umd.edu}
	\affiliation{Condensed Matter Theory Center and Joint Quantum Institute, Department of Physics, University of Maryland, College Park, Maryland 20742, USA}
	
	\author{Zhentao~Wang~\orcidlink{0000-0001-7442-2933}}
	\email{ztwang@zju.edu.cn}
	\affiliation{Center for Correlated Matter and School of Physics, Zhejiang University, Hangzhou 310058, China}
	
	\author{Sankar Das~Sarma~\orcidlink{0000-0002-0439-986X}}
	\affiliation{Condensed Matter Theory Center and Joint Quantum Institute, Department of Physics, University of Maryland, College Park, Maryland 20742, USA}
	
	\date{\today}

	\begin{abstract}
		Motivated by the recently developed classical spin model for Migdal-Eliashberg theory, we develop new numerical and analytical methods based on this spin-chain representation and apply these methods to the Bogoliuov-Tomachov-Morel-Anderson pairing potential, which incorporates the phonon-mediated attraction and Coulomb repulsion. We show that the Monte Carlo method with heat bath updates can efficiently obtain the gap functions even for the situations challenging for the iterative solvers, suggesting an unprecedented robust approach for solving the full nonlinear Migdal-Eliashberg theory. Moreover, we derive the renormalization of all the couplings by tracing out the high-frequency spins in the partition function. The derived analytical renormalization equations produce the well-known $\mu^*$ effect for the Bogoliuov-Tomachov-Morel-Anderson pairing potential and can be generalized to other superconductivity problems. We further point out that several interesting features (e.g., sign changing in the frequency-dependent gap function) can be intuitively understood using the classical spin-chain representation for Migdal-Eliasherg theory. Our results show the advantage of using the spin-chain representation for solving Migdal-Eliashberg theory and provide new ways for tackling general superconductivity problems.
	\end{abstract}
	
	\maketitle
	
	\section{Introduction}
	
	
	Superconductivity is one of the most important quantum phenomena in materials and offers several applications in modern technology. The conventional superconductors can be explained by the phonon-mediated attraction between electrons, i.e., the BCS theory~\cite{BardeenJ1957}. The most successful theory predicting phonon-mediated superconductivity is the Migdal-Eliashberg (ME) theory~\cite{MigdalAB1958,EliashbergGM1960_ME}, which incorporates the retardation effect and the dynamical dependence of the phonon-mediated pairing potential. Remarkably, phonon-mediated superconductivity can persist even if the Coulomb repulsion is stronger than phonon-mediated attraction~\cite{MorelP1962}, resulting in superconductivity from overall repulsive interaction. The phonon-mediated superconductivity applies to a wide range of quantum materials. For example, it has been proposed that phonons can provide sufficiently strong attractive interactions that support observable superconductivity in graphene-based materials~\cite{WuF2018_phonon,WuF2019,LianB2019,WuF2020,SchrodiF2020,SchrodiF2021,LewandowskiC2021,ChouYZ2021_RTG_SC,ChouYZ2022_BBG,ChouYZ2022_ME,VinasBostromE2023}. The phonon-induced superconductivity in graphene \cite{WuF2018_phonon,WuF2019,LianB2019,WuF2020,SchrodiF2020,SchrodiF2021,LewandowskiC2021,ChouYZ2021_RTG_SC,ChouYZ2022_BBG,ChouYZ2022_ME,VinasBostromE2023} would depend crucially on the details of band structures and doping since both the electron-phonon coupling and the Coulomb repulsion would be sensitively dependent on these details, and may even be unknown in the moir\'e structures in general because of strain, relaxation, and twist angle disorder. This may be a possible explanation for why graphene superconductivity is not generically observed in all nominally similar samples as well as why $T_c$ varies from 20mK to 2K depending on the details \cite{CaoY2018,ParkJM2021,HaoZ2021,ParkJM2022,ZhangY2021,BurgGW2022,ZhouH2021,ZhouH2022,ZhangY2023,SuR2023}.

\begin{figure}[t]
		\includegraphics[width=0.4\textwidth]{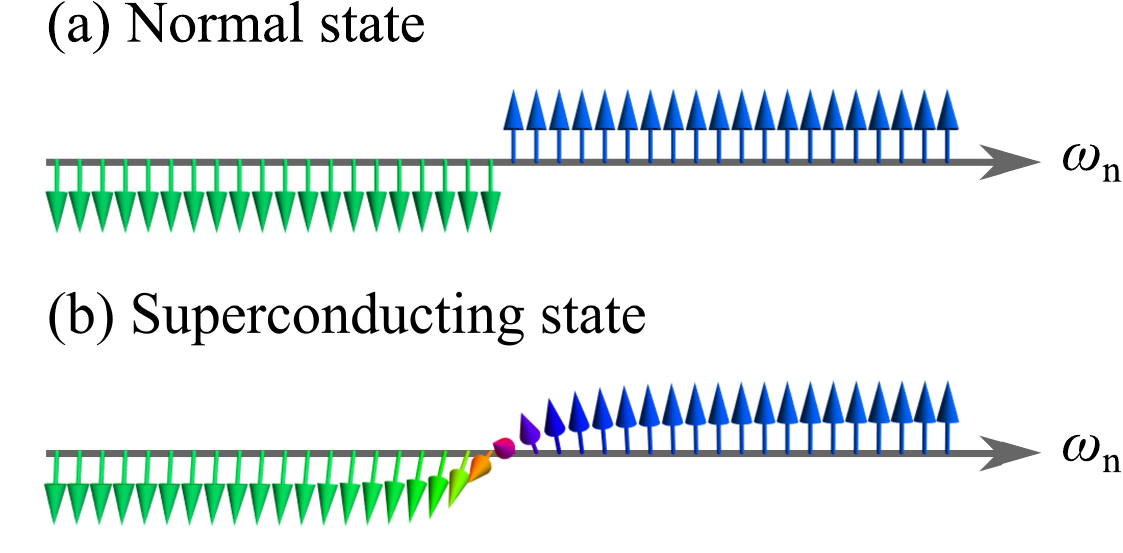}
		\caption{Ground state of Migdal-Eliashberg spin model with phonon-mediated attraction. (a) Normal state spin texture, corresponding to $T>T_c$. (b) Superconducting state spin texture, corresponding to $T<T_c$. The transverse components of spins are related to the anomalous correlation.}
		\label{Fig:Spins}
	\end{figure}

	Recently, Yuzbashyan and Altshuler pointed out that the ME theory can be mapped to a classical spin chain (see Fig.~\ref{Fig:Spins}), bringing new insights to the well-established ME theory~\cite{YuzbashyanEA2022_MESpin,YuzbashyanEA2022_breakdown,YuzbashyanEA2022_SCQC}. Solving the self-consistent Eliashberg equation is equivalent to finding the ground state of the corresponding classical spin chain. This mapping is useful because it suggests that one can study the superconductivity problems with the tools developed for classical spins. In addition, this spin-chain representation may provide a further understanding of results in ME theory. The main goal of this work is to explore possible new methods for the ME classical spin model.

	In this work, we develop new methods for ME spin chain model and apply these methods to the Bogoliuov-Tomachov-Morel-Anderson potential~\cite{BogoljubovNN1958,MorelP1962} (which describes the interplay between phonon-mediated attraction and Coulomb repulsion) as a case study. We show that Monte Carlo (MC) simulations with heat bath updates give accurate solutions even for situations that are difficult for the conventional iterative solvers. We also derive renormalization of couplings for the ME theory by tracing out the high-energy spins. This \textit{spin-decimation} renormalization produces the celebrated $\mu^*$ effect~\cite{MorelP1962} and provides a systematic framework for superconductivity with general pairing potentials. We further point out that several features of superconductivity can be understood intuitively within the spin model. Our results suggest several advantages of using the ME classical spin model.
	
	The rest of the paper is organized as follows: We discuss the mapping between ME theory and the classical spin chain in Sec.~\ref{Sec:Spins}. In Sec.~\ref{Sec:MCHB}, we introduce the MC method with heat bath updates, which efficiently solve the Eliashberg equations. Then, we apply this new numerical solver to the Bogoliuov-Tomachov-Morel-Anderson pairing potential in Sec.~\ref{Sec:BTMA_model}. We further derive the renormalization of couplings in Sec.~\ref{Sec:Ren}. Finally, we discuss several technical issues and future directions in Sec.~\ref{Sec:Discussion}. In the appendices, we discuss the MC implementations with both heat bath and Metropolis updates (Appendix~\ref{App:MC}), iterative solvers (Appendix~\ref{App:It}), and derivation for the spin-decimation renormalization (Appendix~\ref{App:RG}).

	\section{Classical spin model}\label{Sec:Spins}

	We review the mapping between ME theory and classical spin chain following Ref.~\cite{YuzbashyanEA2022_MESpin} in this section. First, we sketch the derivation of ME theory and then discuss the mapping. Then, we discuss the superconducting transition in the classical spin representation.
	
	\subsection{Mapping to classical spins}
	
	To derive the ME theory, we consider electron-phonon models, such as the Holstein model. As shown in Ref.~\cite{YuzbashyanEA2022_MESpin}, the precise form of phonon dispersion or the electron-phonon interaction is not crucial for the derivation. The first step is integrating out the phonons in the imaginary-time path integral formalism, which generates a dynamical fermion-fermion pairing interaction with nontrivial frequency dependence. In the BCS theory, the full dynamical dependence of the pairing interaction is simplified by an effective instantaneous attractive interaction (corresponding to a constant in the frequency domain). Here, we keep the full frequency dependence. Next, we employ the Hubbard-Stratonovich decoupling for the pairing interaction and then integrate out the fermionic fields. After averaging over the spatial degrees of freedom, the ME free energy functional per site is given by~\cite{YuzbashyanEA2022_MESpin}:
	\begin{align}\label{Eq:f_ME}
		\nonumber f_{\text{ME}}=&-2\pi\nu_0T\sum_n\sqrt{(\omega_n+\Sigma_n)^2+|\Phi_n|^2}\\
		&+\nu_0 T^2\sum_{n,m}\left[\Phi^*_{n}(V^{-1})_{n,m}\Phi_m+\Sigma^*_{n}(V^{-1})_{n,m}\Sigma_m\right],
	\end{align} 
	where $\nu_0$ is the density of states, $\omega_n=\pi T(2n+1)$ is the fermionic Matsubara frequency, $n$ is an integer, $T$ denotes the physical temperature, $V^{-1}$ is the inverse pairing interaction, $\Sigma_n$ is a real-valued field associated with the normal electron correlation (e.g., $\langle c^{\dagger}_{\uparrow}c_{\uparrow}\rangle$), and $\Phi$ is a complex-valued field associated with the anomalous electron correlation (e.g., $\langle c_{\downarrow}c_{\uparrow}\rangle$). The ME equations can be derived by minimizing $f_{\text{ME}}$:
	\begin{subequations}\label{Eq:ME_Eqs}
		\begin{align}
			\label{Eq:ME_Eq_Phi}\Phi_n=&\pi T\sum_mV_{n,m}\frac{\Phi_m}{\sqrt{\left(\omega_m+\Sigma_m\right)^2+|\Phi_m|^2}},\\
			\Sigma_n=&\pi T\sum_mV_{n,m}\frac{\omega_m+\Sigma_m}{\sqrt{\left(\omega_m+\Sigma_m\right)^2+|\Phi_m|^2}},
		\end{align}
	\end{subequations}
	where $V$ is the pairing potential. 
	
	As pointed out by Yuzbashyan and Altshuler~\cite{YuzbashyanEA2022_MESpin}, $f_{\text{ME}}$ can be mapped to a spin model. First, we introduce the anomalous ($F_n$) and normal ($G_n$) Green functions as follows:
	\begin{align}
		F_n=&\frac{\Phi_n}{\sqrt{\left(\omega_n+\Sigma_n\right)^2+|\Phi_n|^2}},\\
		G_n=&\frac{\omega_n+\Sigma_n}{\sqrt{\left(\omega_n+\Sigma_n\right)^2+|\Phi_n|^2}}.
	\end{align}
	Since $|F_n|^2+G_n^2=1$, we can parametrize $F_n=S_n^x+iS_n^y$ and $G_n=S_n^z$, where $\vec{S}_n=(S_n^x,S_n^y,S_n^z)$ is a unit vector with three components. Using the spin representation and Eq.~(\ref{Eq:ME_Eqs}), $f_{\text{ME}}$ can be rewritten. One can easily check that
	\begin{align}
		\nonumber&\sum_{n,m}\!\!\left\{\!\Phi^*_{n}\!\left[V^{-\!1}\right]_{n,m}\!\Phi_m\!+\!\Sigma_{n}\!\left[V^{-\!1}\right]_{n,m}\!\Sigma_m\!\right\}
		\\
		=&\pi^2\!\sum_{n,m}V_{n,m}\vec{S}_n\!\cdot\!\vec{S}_m,\\
		\nonumber&\sum_n\!\!\sqrt{\left(\omega_n\!+\!\Sigma_n\right)^2\!+\!|\Phi_n|^2}=\sum_n\left[F_n\Phi_n^*+G_n\left(\omega_n+\Sigma_n\right)\right]
		\\
		=&\!\sum_n\omega_nS^z_n\!+\!\pi T\sum_{n,m}V_{n,m}\vec{S}_n\cdot\vec{S}_m.
	\end{align}
	With the above equations, the spin Hamiltonian for ME theory is given by~\cite{YuzbashyanEA2022_MESpin}
	\begin{align}\label{Eq:H_spin}
		\mathcal{H}_{\text{spin}}\equiv\frac{f_{\text{ME}}}{\nu_0 T}=-2\pi\sum_n\omega_n S_n^z-\pi^2T\sum_{n,m}V_{n,m}\vec{S}_n\cdot\vec{S}_m.
	\end{align}
	The partition function of ME theory now becomes $Z=\int D[\Sigma,\Phi,\Phi^*]e^{-H_{\text{spin}}/\delta}$, where $\delta$ is the level spacing acting as an effective temperature for the spin model (but \textit{not} the physical temperature $T$). With $\delta\rightarrow 0$, solving the Eliashberg equations [Eq.~(\ref{Eq:ME_Eqs})] becomes equivalent to finding the ground state of $H_{\text{spin}}$.
	The first term of $H_{\text{spin}}$ is a site-dependent Zeeman field; the second term describes a long-range classical Heisenberg interaction. $V_{n,m}$ depends on specific model interaction. 
	
	Here, we discuss the procedures of extracting the superconducting gap function via classical spins. First, the gap function is defined by $\Delta(\omega_n)\equiv \Delta_n=\Phi_n/Z_n$, where $Z_n=1+\Sigma_n/\omega_n$. With classical spins, we can parameterize $\vec{S}_n=(\sin{\theta_n}\cos\phi_n,\sin\theta_n\sin\phi_n,\cos\theta_n)$. The amplitude of gap function is $|\Delta(\omega_n)|=|\Phi_n|/Z_n=|\omega_n\tan\theta_n|$, and $\arg\Delta(\omega_n)=\sgn(\omega_n)\phi_n$. In a superconducting state, the spin chain has finite transverse spin components, i.e., $\sin\theta_n\neq 0$. 
	
	\subsection{Superconducting transition in classical spin representation}

	To study the superconducting transition in the classical spin representation, we consider the phonon-mediated attraction $V_{n,m}=g^2/[(\omega_n-\omega_m)^2+\Omega^2]$, where $g$ is the electron-phonon coupling and $\Omega$ is the characteristic phonon frequency. The dimensionless parameter is given by $\lambda=g^2/\Omega^2$. For $\lambda=0$, the ground state of $H_{\text{spin}}$ is described by $\vec{S}_n=\sgn(\omega_n)\hat{z}$ (fully polarized along the local Zeeman field directions) which corresponds to the normal state. 
	The ferromagnetic interaction and the local Zeeman field are compatible for spins with $|n|\gg1$ where the local Zeeman field strongly pin the spins along the $\hat{z}$ ($-\hat{z}$) direction for $n\gg1$ ($n\ll -1$). 
	However, there is frustration between two interactions for spins close to $n=0$: The local Zeeman fields favor $\vec{S}_{1}=-\vec{S}_0=\hat{z}$, while the ferromagnetic interaction tends to align $\vec{S}_0$ and $\vec{S}_1$. The competition between the local Zeeman fields and the ferromagnetic interaction results in a finite transverse component (on the xy-plane), indicating superconductivity.
	
	The competition between the local Zeeman fields ($\omega_n$) and the ferromagnetic interaction ($V_{n,m}$) can be understood intuitively through a minimal two-spin toy model for ME theory, which is described by
	\begin{align}\label{Eq:H_2}
		\mathcal{H}_2=-E_0\left(S_+^{z}-S_-^z\right)-J\vec{S}_+\cdot\vec{S}_-,
	\end{align}
	where $E_0>0$, $J>0$, and $\vec{S}_+$ ($\vec{S}_-$) denotes the spin coupled to the positive (negative) Zeeman field. Without loss of generality, we assume the spins are in the xz plane (i.e., setting the polar angles to zero). The spins can be fully described by the azimuthal angles $\theta_+$ and $\theta_-$. We find that the ground state of $\mathcal{H}_2$ can be described by $\theta_+=\theta$ and $\theta_-=\pi-\theta$, where
	\begin{align}
		\theta=\begin{cases}
			\cos^{-1}\left(\frac{E_0}{2J}\right) &\text{ for } J> E_0/2,\\
			0 &\text{ for } J\le E_0/2.
		\end{cases}
	\end{align}
	This simple model shows two phases -- canted spins with the same transverse component and spins polarized along the local Zeeman fields. The former case corresponds to a superconducting state, while the latter case corresponds to a normal state. Interestingly, this two-spin model predicts superconductivity for $J\ge 2E_0$, a finite critical value of $J$. Connecting to the phonon-mediated pairing potential, $J$ is analogous to the $V_{n,n-1}$, which increases as $T$ decreases. Thus, increasing $J$ is equivalent to decreasing $T$, and this toy model captures the superconducting transition.
	
	\section{Monte Carlo simulation with heat bath updates}\label{Sec:MCHB}

	To obtain the ground state of $\mathcal{H}_{\text{spin}}$ [given by Eq.~(\ref{Eq:H_spin})], we employ the MC method with heat bath updates~\cite{MiyatakeY1986}. We briefly discuss the ideas of heat bath updates in the following, and leave the detailed implementation to Appendix~\ref{App:MC}. 
	
	Consider a single spin-flip update at a given site $i$. The
	probability distribution function can be written as
	\begin{equation}
		p(\theta_{i},\phi_{i})=\frac{1}{Z_{i}}e^{-\beta \mathcal{H}_{i}},
	\end{equation}
	where $\mathcal{H}_{i}=-\vec{H}_{i}^{\text{(eff)}}\cdot\vec{S}_{i}$ is the local Hamiltonian
	that describes the effective Zeeman coupling on site $i$, $\beta=1/\mathcal{T}$
	is an artificial inverse temperature used in the MC simulations (not the physical temperature), and
	$Z_{i}$ is the local partition function associated with $\mathcal{H}_{i}$.
	The source of $\vec{H}_i^\text{(eff)}$ includes both the Zeeman field $2\pi \omega_n \hat{z}$ and the couplings from the neighbors $2\pi^2 T \sum_{m\neq i}V_{i,m} \vec{S}_m$.
	
	It is convenient to work in the local frame where $\vec{H}_{i}^{\text{(eff)}}$
	points to the $+\hat{z}$ direction, with the probability distribution
	of the local coordinates \{$\tilde{\theta}_{i}$, $\tilde{\phi}_{i}$\}
	being
	\begin{equation}
		p(\tilde{\theta}_{i},\tilde{\phi}_{i})=\frac{1}{Z_{i}}e^{\beta H_{i}^{\text{(eff)}}\cos\tilde{\theta}_{i}},
	\end{equation}
	where $H_{i}^{\text{(eff)}}\equiv \left|\vec{H}_{i}^{\text{(eff)}}\right|$.
	
	As a result, we can sample \{$\tilde{\theta}_{i}$, $\tilde{\phi}_{i}$\}
	according to two random numbers \{$r_{1}$, $r_{2}$\} drawn from a uniform
	distribution in the range {[}0, 1{]}:\begin{subequations}\label{eq:R1}
		\begin{align}
			r_{1} & =\int_{0}^{2\pi}\mathrm{d}\phi\int_{0}^{\tilde{\theta}_{i}}\sin\theta\mathrm{d}\theta\,p(\tilde{\theta}_{i},\tilde{\phi}_{i}),\\
			r_{2} & =\tilde{\phi}_{i}/\left(2\pi\right).
		\end{align}
	\end{subequations}
	
	The solutions of Eqs.~(\ref{eq:R1}a) and (\ref{eq:R1}b) are given by~\cite{MiyatakeY1986}:
	\begin{subequations}\label{eq:sol_heatbath}
		\begin{align}
			\cos\tilde{\theta}_{i}&=\frac{1}{\beta H_{i}^{\text{(eff)}}}\ln\left[r_{1}e^{-\beta H_{i}^{\text{(eff)}}}+(1-r_{1})e^{\beta H_{i}^{\text{(eff)}}}\right],\\
			\tilde{\phi}_i &= 2\pi r_2.
		\end{align}
	\end{subequations}

	Finally, we need to perform a rotation to obtain the spin $\vec{S}_{i}$
	in the lab frame:
	\begin{equation}
		R\tilde{\vec{S}}_{i}=\vec{S}_{i},
	\end{equation}
	where $R$ is the rotation matrix such that
	\begin{equation}
		R\hat{z}=\vec{H}_{i}^{\text{(eff)}}/H_{i}^{\text{(eff)}}.
	\end{equation}

	In the simulation, a unit MC sweep consists of $2N$ such heat bath
	updates ($2N$ is the number of sites), each performed on a random site $i$. To obtain the ground
	state, we reduce the artificial MC temperature from sweep-$n$ to sweep-$(n+1)$
	according to
	\begin{equation}
		\mathcal{T}_{n+1}=\alpha \mathcal{T}_{n},
	\end{equation}
	where the parameter $\alpha$ is chosen such that the first 1/4 of
	the total MC sweeps are used to anneal from an initial temperature
	$\mathcal{T}_{0}$ to the target artificial temperature $\mathcal{T}_f$ ($\mathcal{T}_0=0.1\Omega$ and $\mathcal{T}_f=10^{-10}\Omega$ in this work), and the rest 3/4 of the MC sweeps are used
	to further equilibrate the ground state at $\mathcal{T}_f$. For all the cases we have tested in this work, 500 sweeps of heat-bath updates are sufficient for achieving accurate gap functions.
	Note that the typical Metropolis
	updates no longer work at very low temperatures, where the proposed
	new configurations are rejected in most cases, causing the spin configurations
	to be stuck in local minima. In contrast, the heat-bath
	updates have a 100\% acceptance ratio at arbitrarily low temperatures. See Appendix~\ref{App:MC} for a discussion.
	
	In Fig.~\ref{Fig:Tc_ph}, we plot $T_c$ as the function of $\lambda$ and compare the well-known asymptotic formulas~\cite{AllenPB1975,ChubukovAV2020_EM}. The MC simulations with heat bath updates solves the full nonlinear ME equations without linearizing equations. Furthermore, our method is very efficient for general pairing potentials, including the situations that are difficult for the iterative solvers (including the damped scheme~\cite{ChubukovA2019} and moving average, see Appendix~\ref{App:It}). We also find that the heat bath updates are much more efficient than the Metropolis updates (with or without over-relaxation updates \cite{LandauDP2021_book}). See Appendix~\ref{App:MC} for a discussion. Thus, the MC simulations with heat bath updates is suitable for studying complicated pairing potential, such as the $\gamma$ model~\cite{WangY2016,ChubukovAV2020_SCQCP} and the potentials featuring complicated frequency dependence, e.g., Refs.~\cite{ChristensenMH2021,PimenovD2022}.

	\begin{figure}[t]
		\includegraphics[width=0.35\textwidth]{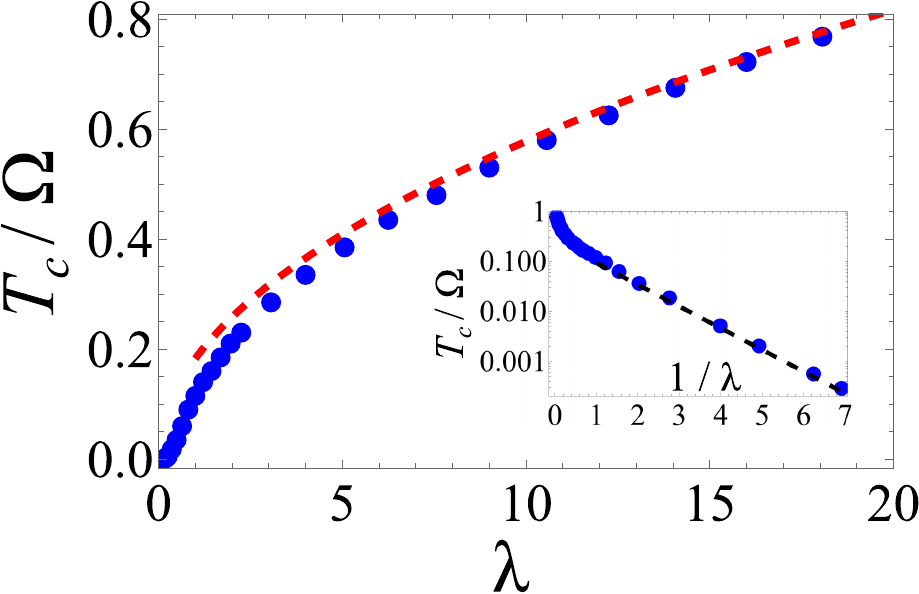}
		\caption{Extracted $T_c$ from Migdal-Eliashberg spin model. The blue dots are obtained using MC simulations with Heat bath updates. The number of spins ($2N$) varies with the temperature, and the frequency cutoff $\Lambda=2\pi T (N+1/2)$ is at least $7\Omega$. The red dashed line denotes the asymptotic formula $T_c/\Omega=0.1827\sqrt{\lambda}$ for $\lambda\gg1$~\cite{AllenPB1975,ChubukovAV2020_EM}. Inset: $T_c$ in the logarithmic scale. The black dashed line denotes the asymptotic formula $T_c/\Omega=0.252e^{-1/\lambda}$ for $\lambda\ll 1$~\cite{ChubukovAV2020_EM}.}
		\label{Fig:Tc_ph}
	\end{figure}

	\section{Application to Bogoliuov-Tomachov-Morel-Anderson pairing potential}\label{Sec:BTMA_model}
	
	In this section, we apply the heat bath MC method to a superconductivity problem with phonon-mediated attraction and Coulomb repulsion. The goal is to demonstrate the advantage of MC simulations with heat bath updates. We also discuss how to understand the results from the ME spin-chain perspective. 
	
	We are interested in the interplay between the phonon-mediated pairing and Coulomb repulsion, which can be described by the Bogoliuov-Tomachov-Morel-Anderson pairing potential given by~\cite{BogoljubovNN1958,MorelP1962,AllenPB1975,ChubukovAV2020_EM,MarsiglioF2020_review}
	\begin{equation}\label{Eq:Vnm_w_Coulomb}
		V_{n,m}=-\mu+\frac{g^2}{(\omega_n-\omega_m)^2+\Omega^2},
	\end{equation}
	where $g$ is the electron-phonon coupling, $\Omega$ is the characteristic phonon frequency, and $\mu>0$ encodes the repulsive instantaneous Coulomb interaction (constant in frequency domain). The dimensionless coupling for phonon-mediated attraction is $\lambda=g^2/\Omega^2$. Note that the form of Eq.~(\ref{Eq:Vnm_w_Coulomb}) is based on the naive continuation of the saddle point equations derived with an attractive pairing interaction. A more careful derivation incorporating the Coulomb repulsion can be found in Ref.~\cite{DalalA2023}. Conventionally, the $-\mu$ term only enters Eq.~(\ref{Eq:ME_Eq_Phi}), corresponding to a constant all-to-all antiferromagnetic transverse spin-spin interaction. We have checked numerically (with iterative solvers) that ignoring the $S^zS^z$ interaction in the $\mu$ term does not change the gap function. This is because the solution always satisfies the favorable spin configuration due to the constant all-to-all $z$-component (Ising-like) spin-spin interaction. In fact, $\sum_nS_n^z=0$ is a manifestation of time-reversal symmetry. Thus, we use the isotropic Heisenberg interaction in the MC simulations with the heat bath updates. 
	
	The existence of superconductivity is determined by $\lambda-\mu^*$~\cite{MorelP1962}, where $\mu^*$ is the renormalized Coulomb repulsion at the energy scale $\sim\Omega$. Intuitively, the Coulomb repulsion in the Cooper channel is marginally irrelevant, resulting in $\mu^*=\mu_0/(1+\mu_0\ln(\Lambda_0/\Lambda^*))$, where $\mu_0$ is the bare Coulomb potential, $\Lambda_0$ is the bare energy cutoff, and $\Lambda^*$ is the cutoff in the renormalized theory. However, the $\mu^*$ effect is not the entire story. The Coulomb repulsion guarantees that the gap function changes sign (apart from the overall phase) at some frequency~\cite{MorelP1962,ColemanP2015_book,PimenovD2022}.

	In the spin model, the Coulomb repulsion corresponds to a constant all-to-all antiferromagnetic Heisenberg interaction. Such an interaction can be rewritten as $\mathcal{H}_{\mu}=\mu|\sum_n\vec{S}_n|^2$, which tends to minimize $|\sum_n\vec{S}_n|$. $\sum_nS^z_n$ is generically zero because the local Zeeman fields in $\mathcal{H}_{\text{spin}}$ favor configurations with $\sum_nS_n^z=0$, suggesting the irrelevance of the $z$-component of the constant all-to-all antiferromagnetic spin-spin interaction. There is a competition between the phonon-mediated attraction (power-law ferromagnetic interaction) and the instantaneous Coulomb repulsion (constant all-to-all antiferromagnetic interaction). A possible solution is to form domains with opposite directions of the transverse components, corresponding to the well-established sign changing of the gap function~\cite{MorelP1962}. In this case, the transverse spins mostly align (due to ferromagnetic interaction) except for those spins near the domain walls, and the total transverse spin components are reduced due to the constant all-to-all antiferromagnetic interaction.
	Analyzing ME theory in terms of classical spins provides an intuitive understanding of the well-known sign changing of gap function in the presence of repulsion.

\begin{figure}[t]
		\includegraphics[width=0.35\textwidth]{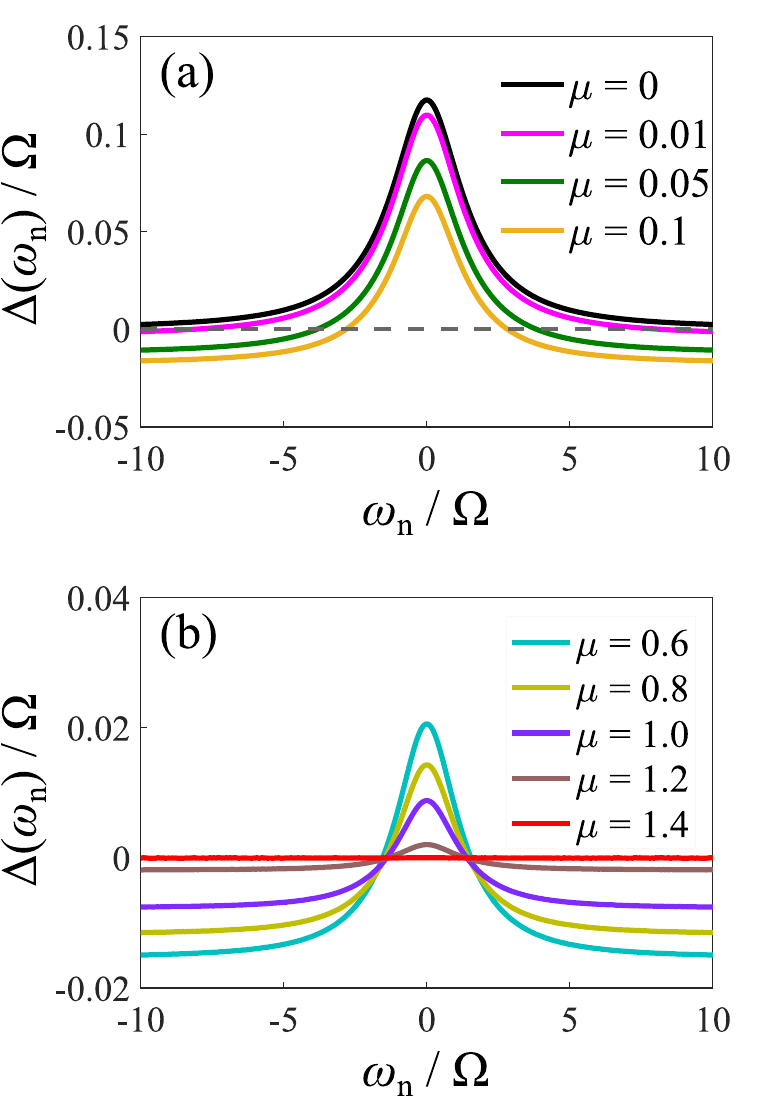}
		\caption{Frequency-dependent order parameter. We plot $\Delta(\omega_n)$ for $\lambda=0.64$, $T=0.01\Omega$, $N=2000$ (i.e., $\Lambda=125.7\Omega$), and several values of $\mu$. (a) Order parameters with $\mu=$0, 0.01, 0.05, and 0.1. The gap function changes sign for $\mu>0$, and $\tilde\omega$ (frequency corresponding to the zero in $\Delta$) shrinks as $\mu$ increases. (b) Order parameters with $\mu=$0.6, 0.8, 1, 1.2, 1.4. All the results are obtained by the MC simulation with heat bath updates.}
		\label{Fig:Delta}
	\end{figure}

	In Fig.~\ref{Fig:Delta}, we compute the superconducting gap using heat bath MC method with $\lambda=g^2/\Omega^2=0.64$, $T=0.01\Omega$, and various representative values of $\mu$. Choosing a different $T$ does not change the qualitative results as long as $T$ is below the transition temperature with $\mu=0$. The gap functions exhibit zeros (denoted by $\tilde\omega$) accompanied by sign changing. The results show superconductivity for $\mu<\mu_c=1.4$, which is much larger than $\lambda=0.64$ used in the calculations. Using the ME spin-chain model, we explicitly establish the well-known result of phonon-mediated superconductivity \cite{MorelP1962}: There is a threshold value of $\lambda$, for a given $\mu$, below which superconductivity is absent. The reverse is also true that, for a given $\lambda$, there is a threshold value of $\mu$ above which superconductivity vanishes.
	Note that the critical value $\mu_c$ here corresponds to $T=0.01\Omega$ rather than for $T=0$. As a result, we obtain $\mu_c^*\approx 0.18$, which is much smaller than $\lambda=0.64$. 
	
\begin{figure}[t]
		\includegraphics[width=0.4\textwidth]{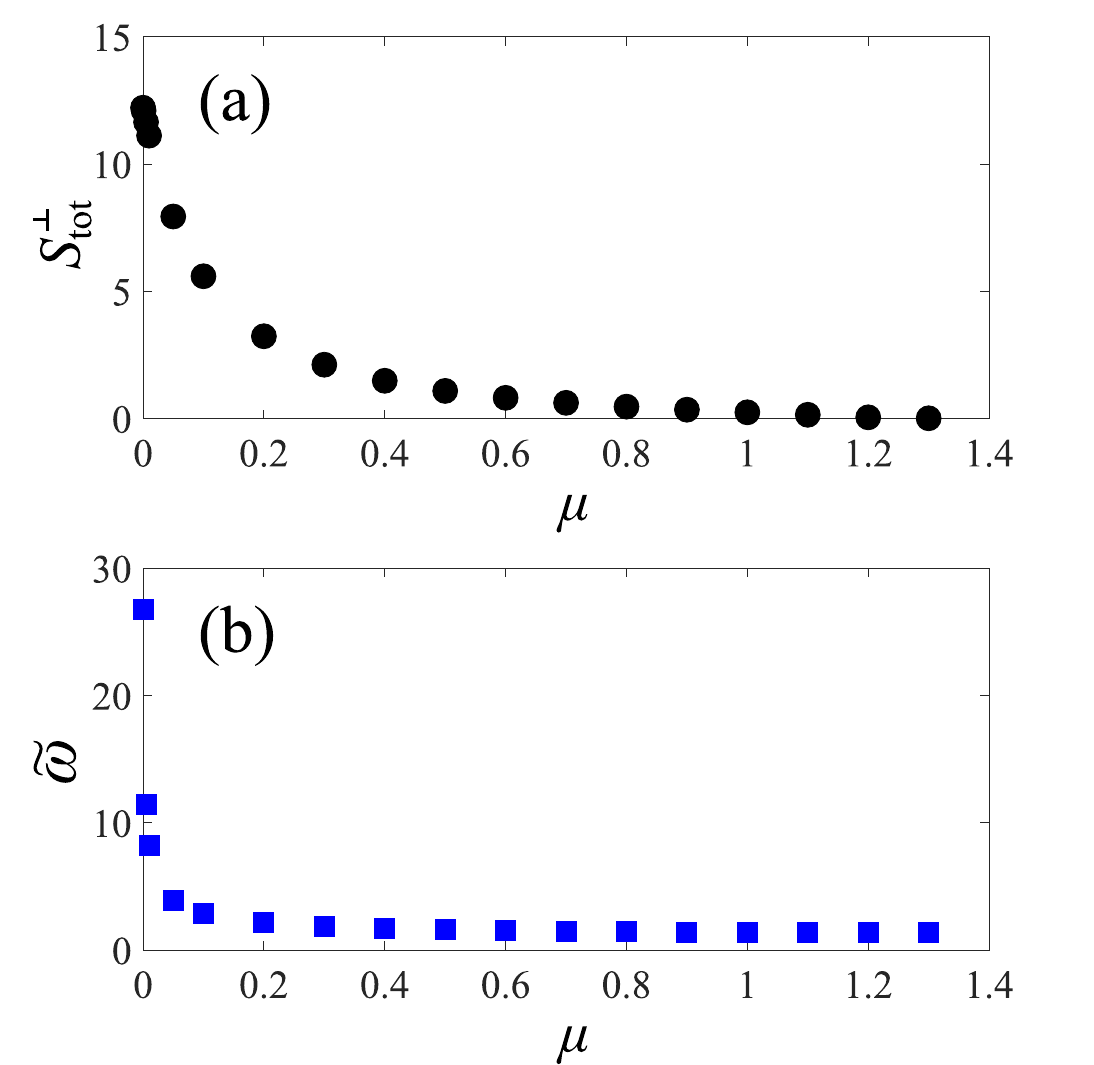}
		\caption{Fine features of order parameter. (a) The total transverse spin component, $S^{\perp}_{\text{tot}}\equiv|\sum_n\vec{S}_n^{\perp}|$, as a function of $\mu$. (b) The frequency associated with the zero in the order parameters, $\tilde{\omega}$, as a function of $\mu$. (Only the positive zero is shown.) Superconductivity is full suppressed at $\mu=1.4$. All the results are obtained by the MC simulation with heat bath updates.
		}
		\label{Fig:Zeros}
	\end{figure}

	As discussed previously, the competition between phonon-mediated attraction (power-law ferromagnetic interaction) and Coulomb repulsion (constant all-to-all antiferromagnetic interaction) can be examined through the total transverse amplitude $S^{\perp}_{\text{tot}}\equiv|\sum_n\left(S_n^x\hat{x}+S_n^y\hat{y}\right)|$. In Fig.~\ref{Fig:Zeros}(a), we show that $S^{\perp}_{\text{tot}}$ decreases as $\mu$ increases, consistent with our intuition that a large $\mu$ favors $S^{\perp}_{\text{tot}}=0$. Note that $S^{\perp}_{\text{tot}}=0$ does not necessarily mean $\sum_n\Delta_n=0$. We also study the evolution of the frequency, $\tilde{\omega}$, associated with the zero in $\Delta_n$. In Fig.~\ref{Fig:Zeros}(b), we find that the $|\tilde{\omega}|$ decreases and then converges to a small frequency ($2\pi T\times 20.5=1.3509\Omega$) as $\mu$ increases to $\mu_c$. The qualitative trend of $\tilde{\omega}$ is similar to the zero-temperature study in Ref.~\cite{PimenovD2022} except that $\tilde{\omega}$ does not approach zero in our finite-temperature calculations.
	
	\section{Renormalization of spin model}\label{Sec:Ren}

	The spin-chain representation of ME theory allows for an explicit derivation of renormalization of parameters. In the spin model, the positions of the spins represent Matsubara frequencies. Thus, integrating out the high frequencies is equivalent to tracing out the classical Heisenberg spins at the boundaries. In the following, we show renormalization for $\mathcal{H}_{\text{spin}}$ by decimating of the boundary high-frequency spins, which can be done analytically within some approximations.
	
	First, we discuss the decimation procedure for one spin. For a spin at site $n$, the equation of motion is governed by the effective Hamiltonian $\mathcal{H}_n=-\vec{H}_n\cdot\vec{S}_n$, 
	where 
	\begin{align}
		\vec{H}_n=2\pi\omega_n\hat{z}+2\pi^2 T\sum_{m\neq n}V_{n,m}\vec{S}_m.
	\end{align}
	In the partition function, we can integrate out the site $n$ and obtain the correction to spin Hamiltonian~\cite{FisherME1964}, 
	\begin{align}
		\delta \mathcal{H}=-\beta^{-1}\ln\sinh(\beta|\vec{H}_n|)+\beta^{-1}\ln(\beta |\vec{H}_n|),
	\end{align}
	where $\beta=\delta^{-1}$ is the effective inverse temperature of the classical spin chain (not the inverse physical temperature), and $\delta$ is the level spacing of the physical system. After taking $\beta\rightarrow \infty$, 
	$\delta \mathcal{H}= -|\vec{H}_n|$.
	In most cases, the local Zeeman field term ($2\pi\omega_n\hat{z}$) of $\vec{H}_n$ dominates. Thus, we can derive the correction to the Hamiltonian, ignoring the $O(|\omega_n|^{-2})$ contributions. See Appendix~\ref{App:RG} for derivations.

	\begin{figure*}[t]
		\includegraphics[width=0.8\textwidth]{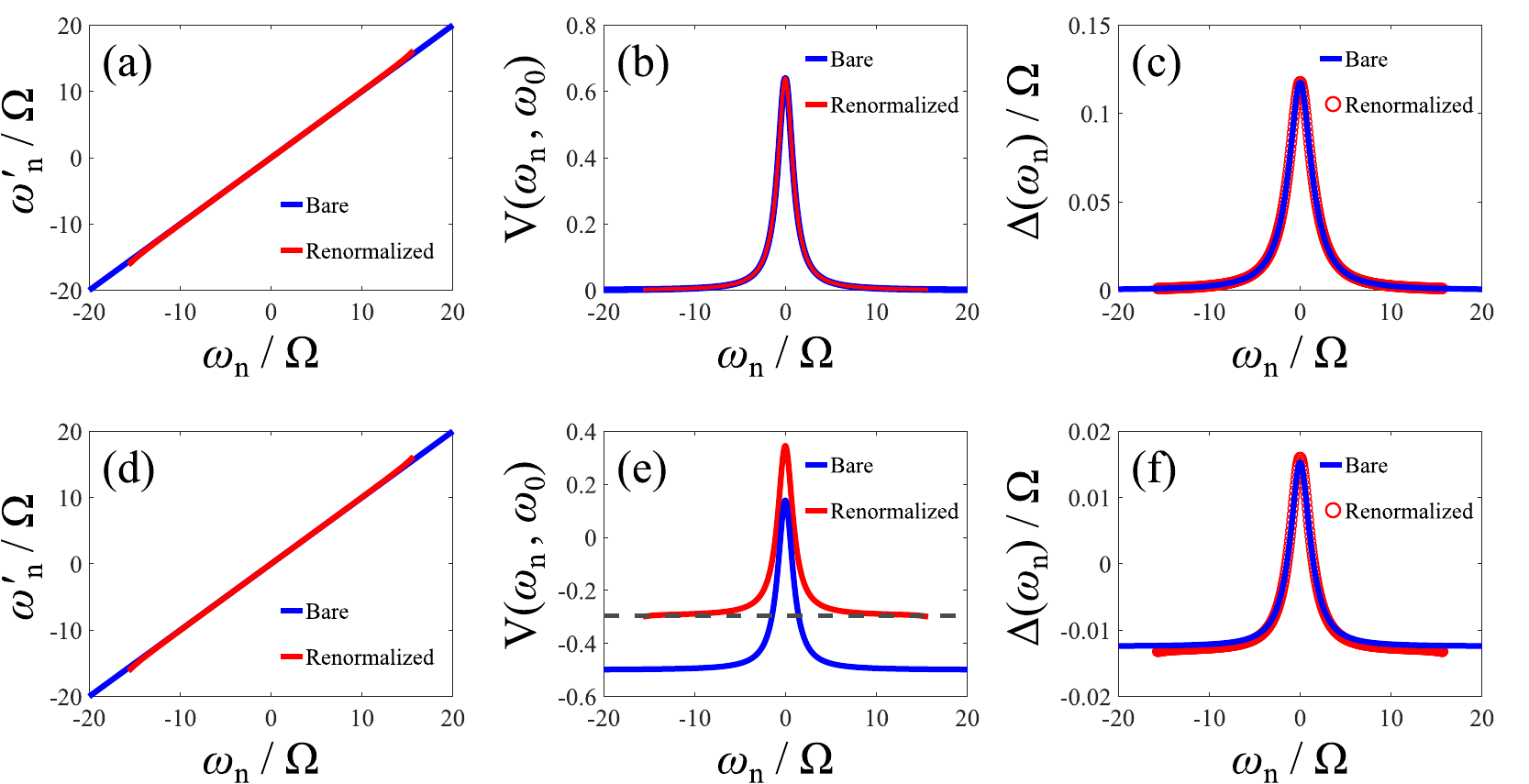}
		\caption{Renormalized couplings from decimation of spins based on Eq.~(\ref{Eq:dH_after_decimation}). $\lambda=0.64$ and $T=0.01\Omega$ are used in these plots. $\mu=0$ is used in (a)-(c), and $\mu=0.5$ is used in (d)-(f). We integrate out the high-energy spins with $2\pi T (N_0-1/2)<|\omega_n|\le 2\pi T(N_i-1/2)$, where $N_0=250$ and $N_i=1000$. (a), (d) Comparison between the bare and renormalized frequencies.  (b), (e) Comparison between bare and renormalized pairing interactions. The black dashed line in (e) represents the Coulomb potential estimated by $-\mu^*=-0.5/(1+0.5\ln 4)=-0.2953$. (c), (f) Comparison between solutions from bare and renormalized couplings. The blue lines are obtained by solving the Eliashberg equation with the bare interaction and $N=1000$; the red circles are obtained by solving the renormalized Eliashberg equation with the renormalized interaction and $N=250$.
		}
		\label{Fig:Ren}
	\end{figure*}
	
	After tracing out the boundary spins (i.e., $n=-N$ and $n=N-1$ sites), the corrections to $H_{\text{spin}}$ are described by:
	\begin{align}
		\nonumber\delta \mathcal{H}_{\text{spin}}\!=\!&-\!2\pi^2T\sum_{n}\left(V_{n,N-1}^z-V_{n,-N}^z\right)S_n^z\\
		\label{Eq:dH_after_decimation}&-\!\frac{\pi^3T^2}{\Lambda}\!\sum_{n,m}\!\left(V_{n,N-1}^{\perp}V_{m,N-1}^{\perp}\!+\!V_{n,-N}^{\perp}V_{m,-N}^{\perp}\right)\!\vec{S}_n^{\perp}\!\cdot\!\vec{S}_m^{\perp},
	\end{align}
	where the sites $n,m\in\{-N+1,\dots,N-2\}$, $\vec{S}_n^{\perp}\equiv S_n^x\hat{x}+S_n^y\hat{y}$, and $\Lambda=|\omega_{-N}|=\omega_{N-1}$ is the original frequency cutoff. The superscripts $z$ and $\perp$ in $V$ denote the interaction of z and transverse components, respectively. Before tracing out spins, the spin-spin interaction is isotropic, i.e., $V_{n,m}^z=V_{n,m}^{\perp}$. After tracing out the spins, $V_{n,m}^z$ and $V_{n,m}^{\perp}$ become unequal, and single-ion anisotropy terms [e.g., $D(S_n^z)^2$] are created. We have checked numerically that the single-ion anisotropy terms generate $O(\Lambda^{-2})$ correction, which we ignore in this work. Equation~(\ref{Eq:dH_after_decimation}) applies for general pairing potentials (as long as $|\omega_n|\gg |V_{n,m}|$) and indicates a systematic procedure of reducing the number of Matsubara frequencies in solving Eliashberg equations. 
	
	In Fig.~\ref{Fig:Ren}, we apply the analytical results to the Bogoliuov-Tomachov-Morel-Anderson pairing potential [Eq.~(\ref{Eq:Vnm_w_Coulomb})] and plot the renormalized parameters as well as the gap functions using the renormalized equations through iteration. The results suggest that the low-frequency gap functions can be reliably extracted using the derived renormalized parameters with a reduced number of sites (frequencies), implying a reduction of computational complexity for ME theory. The spin-decimation renormalization is similar to the folding procedure in solving Eliashberg equations for phonon-mediated superconductivity~\cite{MarsiglioF2020_review} except that renormalization of $\omega_n$ is also taken into account (albeit small compared to $\omega_n$ in this case) in our spin-decimation renormalization procedure. 
	
	To connect our results to the renormalization group equations, we consider $T\rightarrow 0$ and derive the corresponding flow equations as follows:
	\begin{subequations}\label{Eq:Scaling}
		\begin{align}
			\frac{d\omega_n}{d\Lambda}=&-\!\frac{1}{2}\left(V_{n,N-1}^{z}-V_{n,-N}^{z}\right),\\
			\frac{d V_{n,m}^{\perp}}{d\Lambda}=&-\frac{1}{2\Lambda}\left(V^{\perp}_{n,N-1}V^{\perp}_{m,N-1}+V^{\perp}_{n,-N}V^{\perp}_{m,-N}\right),
		\end{align}
	\end{subequations}
	where we have used $d\Lambda=2\pi T$ with $T\rightarrow 0$.
	Note that these flow equations are of the poor man's scaling type, i.e., the cutoff is not rescaled. We emphasize that Eq.~(\ref{Eq:Scaling}) is general and applies to pairings beyond the phonon-mediated superconductivity.

	Now, we examine the Bogoliuov-Tomachov-Morel-Anderson pairing potential. The right-hand side of Eq.~(\ref{Eq:Scaling}a) is proportional to $\Lambda^{-3}$ for phonon-mediated attraction, indicating a very weak renormalization in $\omega_n$ as shown in Fig.~\ref{Fig:Ren}(a) and (d). The result validates the absence of renormalization in $\omega_n$ in the conventional folding treatment for phonon-mediated superconductivity~\cite{MarsiglioF2020_review}. For $|n|,|m|\ll N$ and $|\omega_n-\omega_m|\gg\Omega$, equation~(\ref{Eq:Scaling}b) is reduced to $d\mu/d\Lambda=\mu^2/\Lambda$, reproducing marginally irrelevant flow and the famous $\mu^*$ formula~\cite{MorelP1962}, $\mu^*=\mu/(1+\mu\log(\Lambda/\Lambda^*))$, where $\Lambda^*$ and $\mu^*$ denote the new cutoff and the corresponding renormalized Coulomb pseudopotential. In fact, the renormalization of the dynamical phonon-mediated attraction is quite weak as shown in Fig.~\ref{Fig:Ren}, where the primary renormalization is the $\mu^*$ effect. Moreover, the renormalization of $\mu$ in Figure \ref{Fig:Ren}(e) agrees with the $\mu^*$ formula. The above findings indicate that the spin-decimation renormalization procedure not only captures the $\mu^*$ effect but also provides a systematic framework for flows of coupling constants in general pairing Hamiltonian, including more complicated potentials, e.g., the $\gamma$ model~\cite{WangY2016,ChubukovAV2020_SCQCP}. For example, we expect nontrivial renormalization in $\omega_n$ and $V_{n,m}$ for a sufficiently small exponent $\gamma$ in the $\gamma$ model.

	\section{Discussion}\label{Sec:Discussion}
	
	We have shown two new approaches for studying the ME theory with classical spin-chain representation. First, we point out that the MC simulation with heat bath updates is efficient and reliable in obtaining solutions for all the parameters we have considered in this work. This new MC solver for ME theory
	can	outperform the iterative solvers at large system sizes and with complicated pairing potentials. 
	We have also derived renormalization of the couplings with the decimation of high-frequency spins, providing a systematic framework 
	that reproduces the $\mu^*$ effect~\cite{MorelP1962}
	for the Bogoliuov-Tomachov-Morel-Anderson pairing potential~\cite{BogoljubovNN1958,MorelP1962}. 
	Remarkably, both methods are not limited to the specific pairing potential studied in this work, but are applicable to general superconductivity problems with arbitrary pairing potentials.
	
	Besides the results summarized above, we also emphasize several new insights from the classical spin representation of ME theory. First, we have constructed a minimal two-spin model capturing the superconducting transition, providing an intuitive way to understand the superconducting transition. Second, the sign changing of the gap function in the presence of Coulomb repulsion can be easily understood in the spin model because the constant all-to-all antiferromagnetic interaction from Coulomb repulsion tends to minimize the total transverse spin components. The sign changing is not immediately transparent in the self-consistent nonlinear Eliashberg equations but is apparent in the classical spin representation. Third, the spin model provides a natural way to understand the irrelevance of the constant all-to-all antiferromagnetic Ising spin-spin interaction (i.e., $\sum_{n,m}\mu S_n^zS_m^z$) in the problem with the Bogoliuov-Tomachov-Morel-Anderson pairing potential. Last, the renormalization group equations [Eq.~(\ref{Eq:Scaling})] from spin decimation can be used for the search for unconventional superconductivity, showing another advantage of using the classical spin model.

	Now, we discuss several limitations in the numerical and analytical methods. First, the heat bath update described in this paper cannot be directly used for Hamiltonians that include single-ion anisotropy terms, e.g., $D(S_n^z)^2$. In such a case, the heat bath method needs complicated modifications to numerically sample the local probability distribution for \{$\theta_i$, $\phi_i$\} (see Appendix~\ref{App:MC} for a discussion). This is why we incorporate the Coulomb repulsion as a Heisenberg interaction rather than an XX interaction (as commonly seen in literature~\cite{AllenPB1975,MarsiglioF2020_review}). Note that the gap function is unaffected by a constant antiferromagnetic all-to-all $S^zS^z$ interaction for the Coulomb repulsion. 
	Another issue is that the spacing of Matsubara frequencies scales as $T$, suggesting that it is difficult to tackle the problems numerically in the zero-temperature limit with a finite frequency cutoff. With respect to the spin-decimation renormalization, we assume that the local Zeeman field term (i.e., $2\pi\omega_n$) is much stronger than the rest of the terms, and the $O(\Lambda^{-2})$ contributions are ignored. Thus, equations (\ref{Eq:dH_after_decimation}) and (\ref{Eq:Scaling}) might acquire corrections when $\Lambda$ is sufficiently small.

	We conclude by discussing several interesting future directions. It is desirable to develop an efficient MC algorithm for ME spin model that is compatible with the single-ion anisotropy term in the spins. A good candidate is the event-chain MC algorithm~\cite{BernardEP2009,KrauthW2021_review}, which has demonstrated high efficiency in several classical spin problems~\cite{MichelM2015,NishikawaY2015}. In this work, we focus only on the solution of even-frequency superconductivity. It might be interesting to explore the odd-frequency superconductivity~\cite{LinderJ2019_RMP} with spin-chain-based methods. Finally, the idea of mapping the saddle point equations to classical Hamiltonian might be applicable to other many-body problems.

	\begin{acknowledgments}
		\textit{Acknowledgments.---} Y.-Z.C. thanks Jay D. Sau and Andrey Grankin for useful discussions.
		This work is supported by the Laboratory for Physical Sciences (Y.-Z.C. and S.D.S.) and the National Natural Science Foundation of China Grant No.~12374124 (Z.W.).
	\end{acknowledgments}	
	
	\appendix
	
	\section{Monte Carlo simulations}\label{App:MC}

	\subsection{Heat bath update}
	A pseudocode of the heat bath update is shown in Algorithm~\ref{alg:heatbath}, which mainly follows Ref.~\cite{MiyatakeY1986}. As we have shown in this paper, the heat bath method is highly efficient in obtaining the ground state of the classical spin Hamiltonian, which has a 100\% acceptance ratio at arbitrarily low temperature (the fake temperature $\mathcal{T}$ in MC simulations). In contrast, the conventional Metropolis update has practically zero acceptance ratio at low temperatures, which significantly slows down the evolution of the spin configurations towards the ground state.
	
	We note that, the heat bath method is not without limitations, especially when the local Hamiltonian includes terms beyond the local effective Zeeman-field description.
	For example, when the full Hamiltonian includes single-ion anisotropy, then the local Hamiltonian should be written as
	\begin{equation}\label{eq:single-ion}
		\mathcal{H}_i = -\vec{H}_i^\text{(eff)} \cdot \vec{S}_i + \vec{S}_i \cdot \left( A_i \vec{S}_i \right),
	\end{equation}
	where $A_i$ is a $3\times 3$ symmetric matrix. When $A_i\propto \text{Diag}(1,1,0)$, a local XX interaction is realized, which is relevant to the interaction terms generated under the spin-decimation renormalization.
	
	In principle, we can still design the heat bath update in the presence of a nonzero $A_i$ matrix. However, there is no longer a simple frame rotation, as mentioned in Sec.~\ref{Sec:MCHB}, that could help us complete the integrals for the cumulative marginal distribution functions for \{$\theta_i$, $\phi_i$\} in analytical form. As a result, the sampling of \{$\theta_i$, $\phi_i$\} has to be done numerically. 
	
	One way is to perform numerical integration. Then we can use bisection method to find \{$\theta_i$, $\phi_i$\} for given random numbers \{$r_1$, $r_2$\}, since the cumulative distribution function is monotonic in its arguments. This would require a lot of computation of the numerical integration, where a high precision is necessary to maintain the monotonicity. For a number of $n$ integration steps, the heat bath update is a factor of $n$ slower. In this case, a highly optimized code is required to reach the ground state with realistic time cost.
	
	\begin{algorithm}[!tbp]
		\caption{Heat bath update}\label{alg:heatbath}
		\For{$\text{sweep} = 1,2,\ldots, \text{sweep}_\text{max}$}{
			\tcc{a unit sweep comprises $N_\text{sites}$ updates}
			\For{$\text{step}=1,2,\ldots,N_\text{sites}$}{
				draw \{$r_1$,$r_2$\} from uniform distribution in $[0,1]$\;
				\BlankLine
				\BlankLine
				
				pick a random site $i$\;
				compute $\vec{H}_i^\text{(eff)}$ and $H_i^\text{(eff)} \equiv |\vec{H}_i^\text{(eff)}|$\;
				\eIf{$H_i^\text{(eff)}==0$}{
					$\hat{H}_i^\text{(eff)} \equiv (0,0,1)$\;
				}{
					$\hat{H}_i^\text{(eff)} \equiv \vec{H}_i^\text{(eff)} / H_i^\text{(eff)}$\;
				}
				\BlankLine
				\BlankLine
				
				\tcc{note: should treat the $\beta H_i^\text{(eff)} \ll 1$ and $\beta H_i^\text{(eff)} \gg 1$ limits carefully}
				compute \{$\tilde{\theta}_i$, $\tilde{\phi}_i$\} according to Eq.~\eqref{eq:sol_heatbath}\;
				$\tilde{\vec{S}}_i = \left( \sin \tilde{\theta}_i \cos \tilde{\phi}_i, \sin \tilde{\theta}_i \sin \tilde{\phi}_i, \cos \tilde{\theta}_i \right) $\;
				
				\BlankLine
				\BlankLine
				\uIf{$\hat{H}_i^\text{(eff)} == \hat{z}$}{
					$\vec{S}_i = \tilde{\vec{S}}_i$\;
				}
				\uElseIf{$\hat{H}_i^\text{(eff)} == -\hat{z}$}{
					$\vec{S}_i = \text{diag}\{1,-1,-1\}\tilde{\vec{S}}_i$\;
				}\Else{
					$\hat{\omega}_i \equiv (-\hat{H}_i^y, \hat{H}_i^x, 0) / \sqrt{(\hat{H}_i^x)^2 + (\hat{H}_i^y)^2}$\;
					$\alpha_i = \arccos (\hat{H}_i^z)$\;
					$\vec{S}_i = R(\hat{\omega}_i, \alpha_i) \tilde{\vec{S}}_i$\; 
					\tcc{$R(\hat{\omega}_i,\alpha_i)$: rotation matrix with axis $\hat{\omega}_i$ and angle $\alpha_i$}
				}
			}
			
			\BlankLine
			change $\mathcal{T}$ if in annealing stage\; 
		}
	\end{algorithm}
	
	Another way is to sample \{$\tilde{\theta}_i$, $\tilde{\phi}_i$\} using an extra layer of Metropolis updates. In other words, given the local probability distribution, a series of Metropolis updates are performed at site $i$ until \{$\tilde{\theta}_i$, $\tilde{\phi}_i$\} satisfy the Boltzmann distribution determined by the Hamiltonian~\eqref{eq:single-ion}. Similarly to the numerical integration scheme, the extra Metropolis update on top of the heat bath method would significantly slow down the code, and a highly optimized code is desired for this method to work.
	
	\subsection{Metropolis update}

	In a typical Metropolis update, one proposes a new spin configuration randomly distributed on the $S^2$ sphere. The acceptance of the update is controlled by a probability $\min(e^{-\Delta E/\mathcal{T}},1)$, where $\Delta E$ is the energy difference and $\mathcal{T}$ is the MC temperature. The Metropolis update can be applied to a general spectra of Hamiltonians, as long as the energy difference $\Delta E$ can be evaluated numerically. However, at small $\mathcal{T}$, most of the proposed states have $\Delta E > 0$, that leads to a negligible acceptance ratio. In other words, the Metropolis update often get stuck at low-$\mathcal{T}$, which is insufficient for the purpose of obtaining the ground state.

\begin{figure}[t]
		\includegraphics[width=0.375\textwidth]{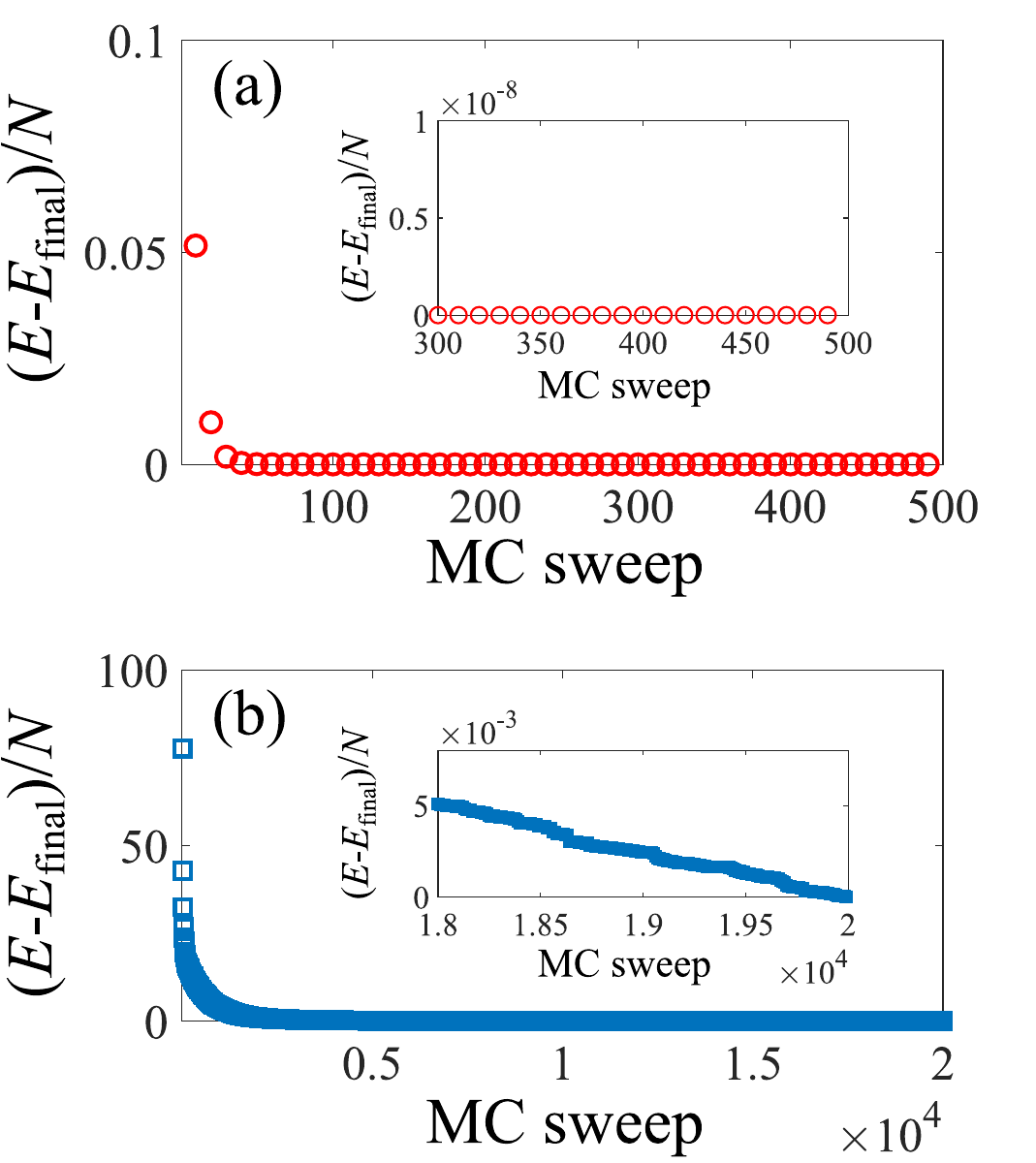}
		\caption{Energy evolution with heat bath and Metropolis updates. We plot $(E-E_{\text{final}})/N$, where $E_{\text{final}}$ is the final energy during the MC simulation. $N=1000$, $g=0.8\Omega$, $\mu=0$, and $T=0.01\Omega$ are used in the MC simulations. (a) MC with heat bath updates. The thermal annealing parameters are $\mathcal{T}_0=0.1\Omega$ and $\mathcal{T}_f=10^{-10}\Omega$. The energy converges after 130 MC sweeps. Inset: The energy evolution between 300 and 500 MC sweeps. (b) MC with Metropolis updates. The thermal annealing parameters are $\mathcal{T}_0=0.1\Omega$ and $\mathcal{T}_f=10^{-3}\Omega$. The energy is evolving during the entire MC simulation. Inset: The energy evolution between 18000 and 20000 MC sweeps. 
		}
		\label{Fig:MC_energy}
	\end{figure}

	We compare the numerical results of heat bath and Metropolis updates in the following. First, we plot the energy evolution as a function of MC sweeps. In Fig.~\ref{Fig:MC_energy}(a), the energy converges after 130 MC sweeps with heat bath updates. In contrast, the MC simulation with Metropolis updates does not achieve convergence of energy even after 20000 MC sweeps, as we show in Fig.~\ref{Fig:MC_energy}, suggesting the inefficiency of Metropolis updates in the ME classical spin model. 
	
\begin{figure}[t]
		\includegraphics[width=0.4\textwidth]{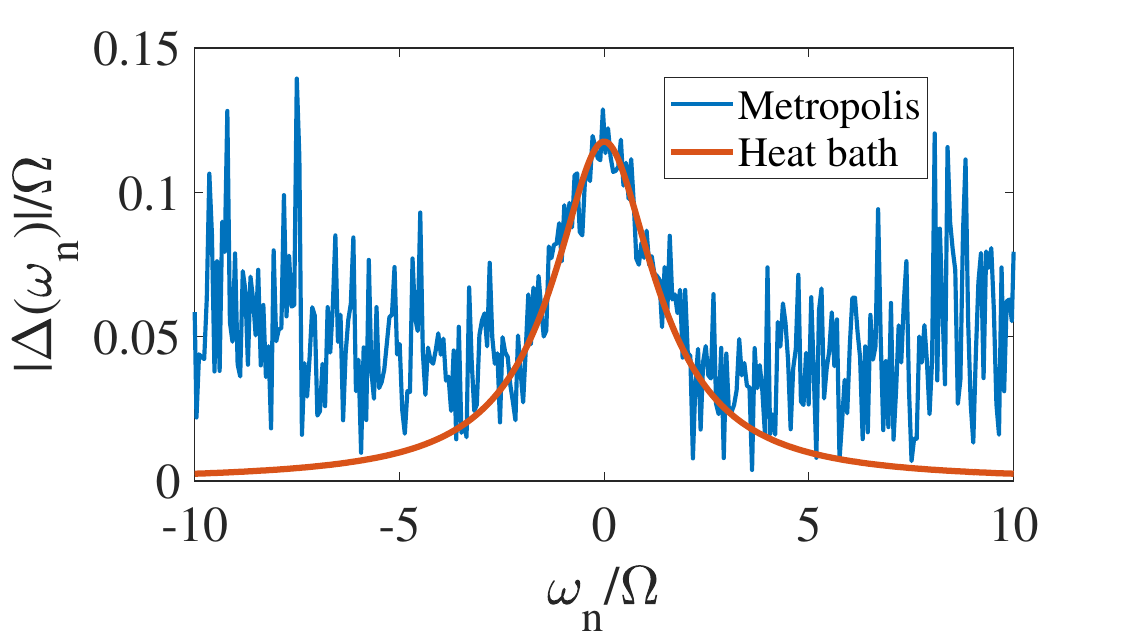}
		\caption{Order parameters from different MC methods. The blue line indicates the result from the MC simulation with Metropolis updates; the red line indicates the result from the MC simulation with heat bath updates. We consider $N=1000$, $g=0.8\Omega$, and $T=0.01\Omega$. $2\times10^4$ MC sweeps are performed for the Metropolis updates. $500$ MC sweeps are performed for the heat bath updates. It is clear that the MC simulation with heat bath updates is significantly better than the MC simulation with Metropolis updates. The heat bath results are consistent with the gap function obtained from iterative solvers.}
		\label{Fig:Delta_comparison}
	\end{figure}

	Next, we examine the states obtained from two methods. In Fig.~\ref{Fig:Delta_comparison}, we plot the order parameters extracted from the heat bath updates after 500 MC sweeps (red line) and the Metropolis updates after 200000 MC sweeps (blue line). It is clear that the MC simulation with heat bath updates gives an accurate solution, while the order parameters extracted from the Metropolis updates are far from being satisfactory. 
	
	There are possibly a few ways to improve the Metropolis method. In fact, we have tried combining the Metropolis updates with over-relaxation~\cite{LandauDP2021_book}, but the results are still much worse than the heat bath updates. Instead of sampling the $S^2$ sphere, one can also design new spin configurations in a small cone whose center overlaps with the current spin direction, then accept/reject using the Metropolis scheme. While such updates should clearly increase the acceptance ratio, the size of the cone must be small enough at low $\mathcal{T}$, and as a result the evolution of the spin configurations would not be very fast.

	\section{Iterative solvers}\label{App:It}
	
	In superconductivity literature, the conventional solvers for the ME equations are based on iterative methods, which can be implemented straightforwardly. In this section, we discuss several variants of iterative solvers and compare the performance with the Bogoliuov-Tomachov-Morel-Anderson pairing potential~\cite{BogoljubovNN1958,MorelP1962}.
	
	\subsection{Standard iteration}
	
	We consider an equation of interest given by $x=f(x)$, where $x$ is a variable and $f(x)$ is some function of $x$. This equation can, in principle, be solved through iteration. Assuming an initial ansatz $x=x_0$, we compute the right-hand side $f(x_0)$ and then choose $x_1=f(x_0)$. Then, we keep iterating the equation with $x_{M+1}=f(x_M)$, where $M$ is a positive integer indicating the iteration time. The iteration stops when $|x_{M+1}-x_{M}|< \epsilon$, where $\epsilon$ is the error.
	
	For ME theory, we define the error,
	\begin{align}\label{Eq:error}
		\epsilon=\sqrt{\sum_{j=-N}^{N-1}\left[\left(\frac{\Phi_j-\Phi_j^{(\text{new})}}{||\Phi_j||}\right)^2+\left(\frac{\Sigma_j-\Sigma_j^{(\text{new})}}{||\Sigma||}\right)^2\right]},
	\end{align}
	where $||A||$ denotes the Euclidean norm of the array $A$.
	
	\subsection{Moving average}
	
	The standard iteration procedure does not guarantee convergence because the result after iteration might overshoot, causing divergence in the iteration. One improved method is using $x_{M+1}=(1-w)x_M+wf(x_M)$ in each update, where $w$ is a weighting parameter between $0$ and $1$. This scheme is in the spirit of the moving average (also known as exponentially weighted average), which effectively averages over the recent $1/w$ outcomes. This scheme tends to smooth out the runaway iteration flows in the conventional iteration scheme. Thus, the moving-average scheme can achieve convergence for most of the cases. However, the number of iterations can be large.
	
	\subsection{``Damping'' scheme}
	
	Another scheme is discussed in Chubukov {\it et al.}~\cite{ChubukovA2019} and is called ``damping'' iteration, which averages all the previous outcomes from the iteration. This can be implemented by using $x_{M+1}=\frac{M}{M+1}x_M+\frac{1}{M+1}f(x_M)$. This scheme can achieve convergence for most of the cases. However, the number of iterations can be extremely large, depending on the problem.

	\begin{figure}[t]
		\includegraphics[width=0.375\textwidth]{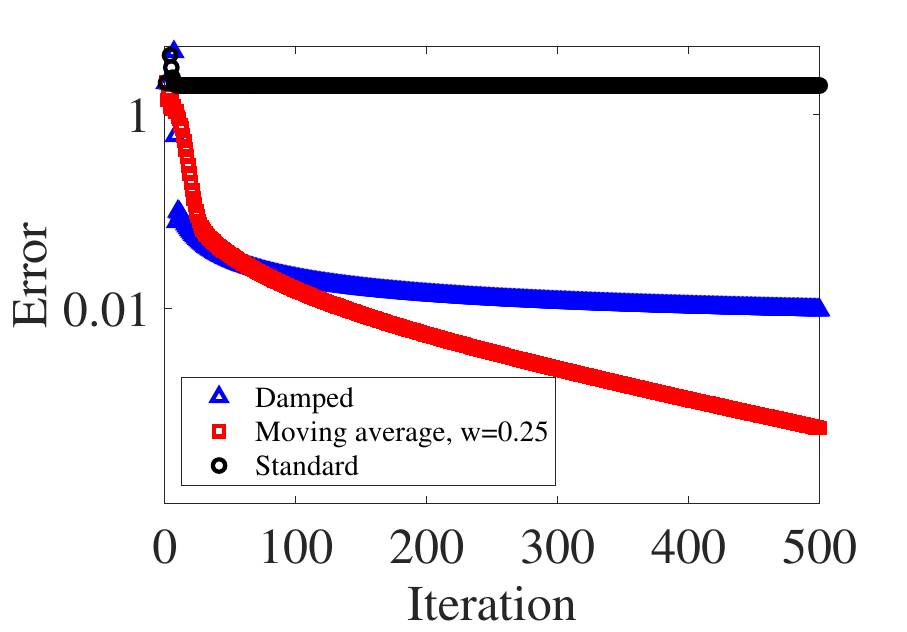}
		\caption{Comparison of errors in different iterative schemes. $N=2000$, $g=0.8\Omega$, $\mu=\Omega$, and $T=0.01\Omega$ are used. The black circles are obtained from the standard iteration; the red squares are obtained from the iteration with moving average (with $w=0.25$, equivalent to averaging over 4 outcomes); the blue triangles are obtained from the iteration with damped scheme, i.e., average over all the previous outcomes.}
		\label{Fig:Iteration_comparison}
	\end{figure}
	
	\subsection{Comparison}
	
	For the easy cases (e.g., $\mu=0$ and $T\ll T_c$), all three iterative solvers can achieve answers with high accuracy, and the standard iterative solver is the most efficient. However, the standard iterative scheme fails to converge for the hard cases (e.g., $\mu>0$ or $T \approx T_c$). In Fig.~\ref{Fig:Iteration_comparison}, we plot the errors as functions of iteration, defined by Eq.~(\ref{Eq:error}), for three different iteration schemes. We find that the standard iteration solver fails to converge, while the moving average and damped scheme gradually achieve convergence as the errors decrease continuously. The results suggest that the standard iterative solver is less reliable than the other two methods, and the moving average scheme is more efficient than the damped scheme. Meanwhile, there is practically no convergence issue in the heat bath MC simulations, and we find that the MC simulation with heat bath updates is much more efficient than all the iterative methods discussed above for the hard cases.
	
	We note that, all the methods considered in this paper, including the heat bath MC method, can in principle produce metastable solutions instead of the true ground states. In all cases, we can start with different initial conditions and test if there are lower energy solutions, which may not always be enough. In this regard, the heat bath MC method is more reliable once we combine it with other standard tricks developed for the classical spin problems. For instance, we have already incorporated the annealing scheme (slowly reducing temperature) in our MC updates so it is less likely to be trapped in local minima. For highly frustrated cases (e.g., spin glass type of interactions), the combination of heat bath, over relaxation, and parallel tempering MC methods is also shown to work~\cite{OgawaT2020}.

	\section{Derivation of spin-decimation renormalization}\label{App:RG}
	
	Now, we provide derivations of the spin-decimation renormalization discussed in Sec.~\ref{Sec:Ren}. 
	We consider the parition function of the classical spin model described by
	\begin{align}
		Z=\left[\prod_{j=-N}^{N-1}\int\frac{d\Omega_i}{4\pi}\right]e^{-\beta \mathcal{H}_{\text{spin}}\left[\vec{S}_{-N},\dots,\vec{S}_{N-1}\right]},
	\end{align}
	where $\Omega_i$ is the solid angle and $\beta$ is the inverse temperature of the classical spin model. (Note that $\beta$ is related to the inverse level spacing as discussed earlier.) Our goal is to integrating out the solid angles $\Omega_{-N}$ and $\Omega_{N-1}$ and derive the renormalized spins chains.
	
	To simplify the calculations, we first consider a toy model of spins given by
	\begin{widetext}
		\begin{align}
			\mathcal{H}_{0}=-\vec{h}\cdot\vec{S}_0-\sum_{n=1}^{M}J_n\vec{S}_0\cdot\vec{S}_n=-\left(\vec{h}+\sum_{n=1}^{M}J_n\vec{S}_n\right)\cdot\vec{S}_0\equiv -\vec{H}_0\cdot\vec{S}_0.
		\end{align}
		Our goal is to derive the effective Hamiltonian after integrating out $\vec{S}_0$. The partition function is given by~\cite{FisherME1964}
		\begin{align}
			Z_0=\int\frac{d\Omega_0}{4\pi}e^{-\beta \mathcal{H}_{0}}=\int\frac{d\Omega_0}{4\pi}e^{\beta \vec{H}_0\cdot\vec{S}_0}=\int_0^{2\pi}\frac{d\phi}{2\pi}\int_{-1}^{1}\frac{d\cos\theta_0}{2}e^{\beta |\vec{H}_0|\cos\theta_0}=\frac{\sinh(\beta|\vec{H}_0|)}{\beta|\vec{H}_0|},
		\end{align}
		where we have set the $z$ axis along the vector $\vec{V}$. The effective Hamiltonian is defined by
		\begin{align}
			\mathcal{H}_{\text{eff}}=&-\beta^{-1}\ln Z_0=-\beta^{-1}\ln\sinh(\beta|\vec{H}_0|)+\beta^{-1}\ln(\beta |\vec{H}_0|).
		\end{align}
		We are interested in the limits $\beta\gg 1$ and $|\vec{h}|\gg \sum_nJ_n$. The effective Hamiltonian becomes
		
		\begin{align}
			\mathcal{H}_{\text{eff}}\approx&-|\vec{H}_0|=-|\vec{h}+\sum_{n}J_n\vec{S}_n|=-h\left(1+h^{-2}\left|\sum_nJ_n\vec{S}_n\right|+2h^{-1}\sum_nJ_n\vec{S}_n\cdot\hat{e}_h\right)^{1/2}\\
			\approx&-h-\sum_nJ_n\vec{S}_n\cdot\hat{e}_h-\frac{1}{2}h^{-1}\left[\left|\sum_nJ_n\vec{S}_n\right|^2-\left(\sum_nJ_n\vec{S}_n\cdot\hat{e}_h\right)^2\right]+O(h^{-2}),
		\end{align}
		where $h\equiv|\vec{h}|$ and $\hat{e}_h$ is the unit vector along $\vec{h}$. We have used $\sqrt{1+x}\approx 1+x/2-x^2/8$ for $x\ll 1$. Without loss of generality, we select $\hat{e}_h=\hat{z}$. The effective Hamiltonian becomes
		\begin{align}
			\mathcal{H}_{\text{eff}}\approx -h-\sum_nJ_n S^z_n-\frac{1}{2h}\sum_{n,n'}J_nJ_{n'}\left(S_n^xS_{n'}^x+S_n^yS_{n'}^y\right).
		\end{align}
		The effective Hamiltonian contribute to a Zeeman field term along $z$ direction and a transverse spin-spin interaction for all the spins coupled to $\vec{S}_0$.
		
		In this work, we consider the ME spin model described by
		\begin{align}
			\mathcal{H}_{\text{spin}}=-2\pi\sum_n\omega_n S_n^z-\pi^2T\sum_{n,m}V_{n,m}\vec{S}_n\cdot\vec{S}_m.
		\end{align}
		For the $n$th spin in the ME spin chain, the dynamics is govern by the Hamiltonian as follows:
		\begin{align}
			\mathcal{H}_{n}=-2\pi\omega_nS_n^z-2\pi^2 T\left[\sum_{m\neq n}V_{mn}\vec{S}_m\right]\cdot\vec{S}_n.
		\end{align}
		Note that there is a factor of 2 in the second term.
		
		Now, we are in the position to derive the correction to $\mathcal{H}_{\text{spin}}$ after integrating out the boundary spins at $-N$ and $N-1$. The partition function is given by
		\begin{align}
			Z=&\left[\prod_{j=-N}^{N-1}\int\frac{d\Omega_i}{4\pi}\right]e^{-\beta \mathcal{H}_{\text{spin}}\left[\vec{S}_{-N},\dots,\vec{S}_{N-1}\right]}\\
			\approx&\left[\prod_{j=-N}^{N-2}\int\frac{d\Omega_i}{4\pi}\right]e^{-\beta \mathcal{H}_{\text{spin}}\left[\vec{S}_{-N},\dots,\vec{S}_{N-1}=0\right]}e^{-\beta\left[-2\pi\omega_{N-1}-2\pi^2 T\sum_{n=-N}^{N-2}V_{n,N-1}S_n^z-\frac{(\pi^2T)^2}{\pi\omega_{N-1}}\sum_{n,m}V_{n,N-1}V_{m,N-1}\vec{S}_n^{\perp}\cdot\vec{S}_m^{\perp}\right]}\\
			\nonumber\approx &\left[\prod_{j=-N+1}^{N-2}\int\frac{d\Omega_i}{4\pi}\right]e^{-\beta \mathcal{H}_{\text{spin}}\left[\vec{S}_{-N}=0,\dots,\vec{S}_{N-1}=0\right]}e^{-\beta\left[-2\pi\omega_{N-1}-2\pi^2 T\sum_{n=-N}^{N-2}V_{n,N-1}S_n^z-\frac{(\pi^2T)^2}{\pi\omega_{N-1}}\sum_{n,m}V_{n,N-1}V_{m,N-1}\vec{S}_n^{\perp}\cdot\vec{S}_m^{\perp}\right]}\\
			&\times e^{-\beta\left[-\left(2\pi|\omega_{-N}|-2\pi^2TV_{N-1,-N}\right)+2\pi^2T\sum_{n=-N+1}^{N-2}V_{-N,n}S^z_n-\frac{(\pi^2 T)^2}{\pi|\omega_{-N}|}\sum_{n,m}V_{n,-N}V_{m,-N}\vec{S}_n^{\perp}\cdot\vec{S}_m^{\perp}+O(|\omega_{-N}|^{-2})\right]}\\
			\propto&\left[\prod_{j=-N+1}^{N-2}\int\frac{d\Omega_i}{4\pi}\right]e^{-\beta \mathcal{H}_{\text{spin}}\left[\vec{S}_{-N}=0,\dots,\vec{S}_{N-1}=0\right]}e^{-\beta\left[-2\pi^2T\sum_n\left(V_{n,N-1}-V_{n,-N}\right)S_n^z-\frac{\pi^4T^2}{\pi\Lambda}\sum_{n,m}\left(V_{n,N-1}V_{m,N-1}+V_{n,-N}V_{m,-N}\right)\vec{S}_n^{\perp}\cdot\vec{S}_m^{\perp}\right]}
		\end{align}
	\end{widetext}
	where we have used $\omega_{N-1}=|\omega_{-N}|=\Lambda$. Several approximations are used in the derivations. First, we drop $O(\Lambda^{-2})$ contributions. Second, we ignore the single-ion anisotropy term, i.e., $D(S_{-N+1}^z)^2$. It is reasonable to omit the $O(\Lambda^{-2})$ contributions as long as $\Lambda$ is much larger than other energy scales. We have checked numerically that the single-ion anisotropy terms give $O(\Lambda^{-2})$ contributions, which we ignore in this work.

	
	%

\end{document}